\documentclass[10pt,pre,twocolumn,showpacs,aps,floats,floatfix,superscriptaddress]{revtex4}

\usepackage{epsfig}
\begin{document}

\newcommand{\avk}{\langle k \rangle}
\newcommand{\fluck}{\langle k^2 \rangle}

\title{The structure of Inter-Urban traffic:\\
A weighted network analysis}

\author{Andrea De Montis}

\affiliation{Dipartimento di Ingegneria del Territorio, Sezione
Costruzioni e Infrastrutture Universit\`a degli Studi di Sassari Via De
Nicola---07100 Sassari, Italy}

\author{Marc Barth\'elemy\footnote{On leave of absence from 
CEA-Centre d'Etudes de Bruy{\`e}res-le-Ch{\^a}tel, 
D\'epartement de Physique Th\'eorique et
Appliqu\'ee BP12, 91680 Bruy\`eres-Le-Ch\^atel, France}}

\affiliation{School of Informatics and Biocomplexity Center,
Indiana University,Eigenmann Hall, 1900 East Tenth Street,
Bloomington, IN 47406}

\author{Alessandro Chessa} 

\affiliation{Dipartimento di Fisica and SLACS,
Universit\`a degli Studi di Cagliari Cittadella Universitaria di
Monserrato, 09042 Monserrato, Italy}

\author{Alessandro Vespignani}

\affiliation{School of Informatics and Biocomplexity Center, 
Indiana University,Eigenmann Hall, 1900 East Tenth Street, 
Bloomington, IN 47406} 
\date{\today} \widetext

\begin{abstract}

We study the structure of the network representing the interurban
commuting traffic of the Sardinia region, Italy, which amounts to 375
municipalities and 1,600,000 inhabitants. We use a weighted network
representation where vertices correspond to towns and the edges to the
actual commuting flows among those. We characterize quantitatively
both the topological and weighted properties of the resulting network.
Interestingly, the statistical properties of commuting traffic exhibit
complex features and non-trivial relations with the underlying
topology. We characterize quantitatively the traffic backbone among
large cities and we give evidences for a very high heterogeneity of
the commuter flows around large cities. We also discuss the interplay
between the topological and dynamical properties of the network as
well as their relation with socio-demographic variables such as
population and monthly income. This analysis may be useful at various
stages in environmental planning and provides analytical tools for a
wide spectrum of applications ranging from impact evaluation to
decision-making and planning support.

\end{abstract}



\maketitle 

\section{Introduction}

The informatics revolution has enabled the systematic gathering and
handling of a large array of large scale networks data sets, enabling
the analysis of their detailed structural features. In particular,
mapping projects of the WWW and the physical Internet offered the
first chance to study the topology and traffic of large-scale
networks. Gradually other studies followed describing population
networks of practical interest in social science, critical
infrastructures and epidemiology (Albert and Barab\`asi, 2002; Newman,
2003; Dorogovtsev and Mendes, 2003; Pastor-Satorras and Vespignani,
2004). The possibility of accessing and mining large-scale data sets
allows a more detailed statistical analysis and theoretical
characterization of correlation patterns, hierarchies and community
structure of these networks, along with the setup of new modeling
frameworks. In particular, the systematic study of these systems has
shown the ubiquitous presence of complex features, mathematically
encoded in statistical distributions with heavy-tails, diverging
fluctuations, self-organization and emerging phenomena and patterns.

The statistical properties induced by these complex features indicate
the presence of topological and structural properties that do not find
an explanation in the paradigm put forward by Erd\"os and Renyi with the
random graph model (Erd\"os and Renyi, 1960). Indeed, even if the
Erd\"os-Renyi model rationalizes the small-world property of networks
(i.e.  the short distance measured in number of links among nodes), it
fails in reproducing the high level of local cohesiveness observed in
many networks (Watts and Strogatz, 1998). Additionally, several of
these networks are characterized by the statistical abundance of nodes
with a very large degree $k$; i.e. the number of connections to other
nodes. For these 'scale-free' networks, the degree probability
distribution $P(k)$ spans over a wide range of $k$ values which signals
the appreciable occurrence of large degree nodes, the 'hubs' of the
system. Finally, it is important to remark that the topological
features of networks turn out to be extremely relevant since they have
a strong impact in assessing their physical properties such as
robustness, vulnerability, or the ability to spread a disease
(Barab\`asi and Albert, 1999; Dorogovtsev and Mendes, 2003;
Pastor-Satorras and Vespignani, 2004).

More recently, the activity on complex networks has been extended to
the characterization of weighted networks. This representation allows
the consideration of features pertaining to the dynamics and traffic
flows occurring on networks, adding another dimension in the
description of these systems. Also in this case, the analysis and the
characterization of weighted quantities have pointed out the presence
of large scale heterogeneity and non-trivial correlations (Barrat et
al, 2004). The weighted graph representation provides valuable
elements that open the path to a series of questions of fundamental
importance in the understanding of networks. Among those, the issue of
how dynamics and structure affect each other and the impact of traffic
flows on the basic properties of spreading and congestion
phenomena. Finally, it spurs the more theoretical questions on how all
these properties may be considered in generative networks models.

In the realm of urban and environmental planning, spatial analysis and
regional science, the interest for complex networks has noticeably
increased during the last years. In his investigation of the structure
of the modern urban fabric, Salingaros (2001 and 2003) invokes the
concepts of small world network and scale free properties and suggests
a parallelism between the city and an ecosystem. Starting from the
principle that connective webs are the main source of urban life,
Salingaros invites planners to apply complex networks analysis to
understand properties such as resilience and self-organization. While
these studies are mostly based on conceptual arguments, other authors
(Shiode and Batty, 2000; Batty, 2001 and 2003) promote the development
of quantitative analyses with a twofold perspective: (i) explaining
the behavior of complex networks such as the world wide web in terms
of geography, demography and economics; and (ii) improving the
efficiency of advanced spatial analysis GIS-based tools, by
integrating the foundation of complex networks analysis into
geographic information science. Batty (2001) along with Salingaros
acknowledges the existence of a link between efficient urban
connectivity and self-preserving behaviors and fosters the development
of complex networks to study the functional mechanisms of cities. It
is interesting to note that a similar approach has been extremely
fruitful in the study of the Internet in relation with the
geographical and social environment (Yook et al, 2002; Gorman, 2001;
Barth\'elemy et al, 2003; Gorman and Kulkarni 2004; Pastor-Satorras
and Vespignani, 2004; Schintler et al, forthcoming).

In the domain of geographic information science, Jiang and Claramunt
(2004) proposed to use complex network tools to investigate
quantitatively the urban space syntax. Aiming at the improvement of
the effectiveness of network GIS advanced analysis, they use
topological measures of connectivity, average path length and
generalised clustering to compare three large street networks of the
cities of G\"avle, Munich and San Francisco. Fertile research
directions are also currently explored in the domain of transportation
systems and human traffic analysis. For example, the small-world
properties of the Boston subway system have been characterized by
Latora and Marchiori (2000-2003) and Sen et al (2003) studied the
topology of the Indian railway network. At the urban level, a network
study has been carried out by Chowell et al (2003) using large scale
simulations (with census and demographic data integration) in order to
describe urban movements of individuals in Portland, Oregon (USA). On
a larger scale, recent studies have focused on the network properties
of major transportation infrastructure such as the US Interstate
highway network and the airport network. (Gastner and Newman, 2004;
Guimera et al, 2003). Finally, Barrat et al (2004) have provided the
first study of the worldwide airports network including the traffic
flows and their correlation with the topological structure. In this
research area, the network approach might provide relevant information
on issues such as traffic analysis and risk assessment in case of
damages and attacks.

In the present work we use a network approach to study the Sardinian
inter-municipal commuting network (SMCN), which describes the everyday
work and study-led movements among $375$ municipalities in the Italian
region of Sardinia. We obtain a weighted network representation in
which the vertices correspond to the Sardinian municipalities and the
valued edges to the amount of commuting traffic among them. We provide
a detailed quantitative study of the resulting network, aiming at a
first characterization of the structure of human traffic at the
inter-city level and its relation with the topological structure
defined by the connectivity pattern among cities. Very interestingly,
while the network topology appears to fit within the standard random
graph paradigm, the weighted description offers a very different
perspective with complex statistical properties of the commuting flows
and highly non-trivial relations between weighted and topological
properties. We discuss these properties in relation with the
geographical, social and demographical aspects of the system and show
the potentiality of the network approach as a valuable tool at various
stages of the policy-making and environmental planning process.

The paper is organized as follows. In the next section, we describe
the dataset. In section 3, we carry out the analysis of the topology
of the network. In section 4, we analyze the traffic and weight
properties of this network and in section 5, we discuss the
implications of our network analysis results in terms of urban and
environmental planning. In section 6, we compare topological and
weighted network analysis results with socio-economic characteristics
of the Sardinian municipalities and we discuss their relations. We
finally summarize our results in section 7, while in section 8 we
discuss the different research directions that this work suggests.

\section{Setting the case study: Data and geographical features}

Sardinia is the second largest Mediterranean island with an area of
approximately $24,000$ square kilometers and $1,600,000$ inhabitants. Its
geographical location and morphological characters have determined an
important history of commercial and cultural relations with
trans-borders external communities and have also favoured the
development of a strong social identity and of important political
movements toward self-reliance and autonomy. At the date of 1991, the
island was partitioned in $375$ municipalities, the second simplest body
in the Italian public administration, each one of those generally
corresponding to a major urban centre (in Figure 1, on the left, we
report the geographical distribution of the municipalities). For the
whole set of municipalities the Italian National Institute of
Statistics (Istat, 1991a) has issued the origin-destination table (ODT)
corresponding to the commuting traffic at the inter-city level. The
ODT is constructed on the output of a survey about commuting behaviors
of Sardinian citizens. This survey refers to the daily movement from
the habitual residence (the origin) to the most frequent place for
work or study (the destination): the data comprise both the
transportation means used and the time usually spent for
displacement. Hence, ODT data give access to the flows of people
regularly commuting among the Sardinian municipalities. In particular
we have considered the external flows $i\to j$ which measure the movements
from any municipality $i$ to the municipality $j$ and we will focus on the
flows of individuals (workers and students) commuting throughout the
set of Sardinian municipalities by all means of transportation. This
data source allows the construction of the Sardinian inter-municipal
commuting network (SMCN) in which each node corresponds to a given
municipality and the links represent the presence of a non-zero flow
of commuters among the corresponding municipalities.
\begin{figure}[ht]
\vspace*{.1cm} 
\centerline{
\includegraphics*[width=0.20\textwidth]{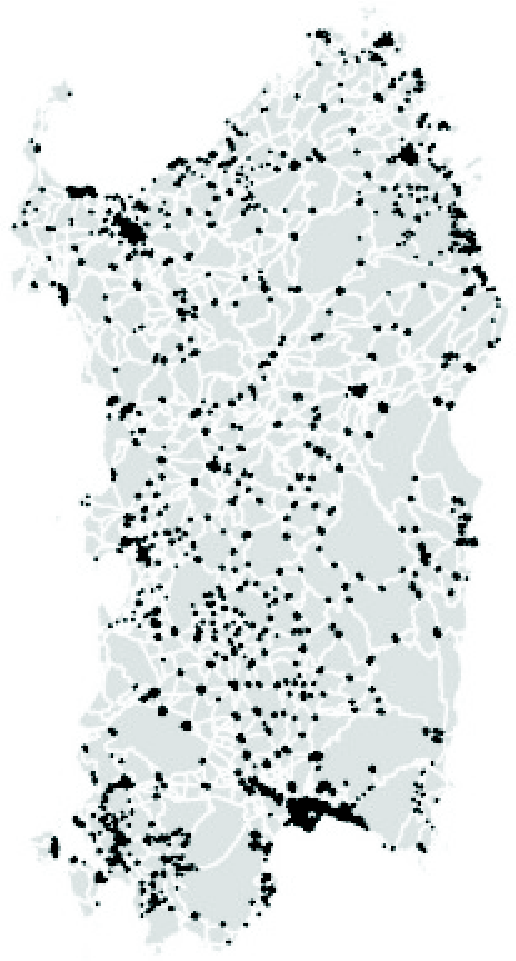}
\includegraphics*[width=0.28\textwidth]{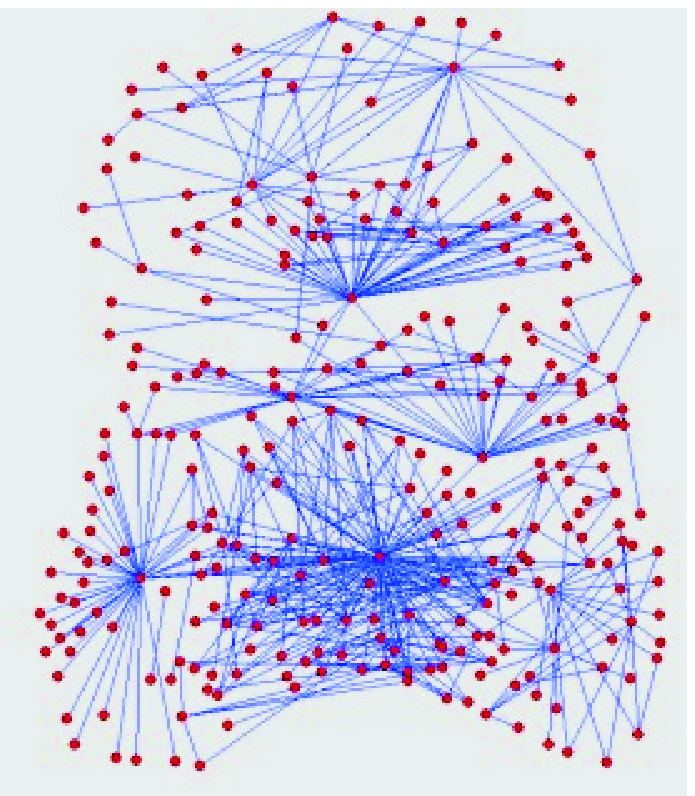}
}
\caption{ Geographical versus topologic representation of the SMCN:
the nodes (red points) correspond to the towns, while the links to a
flow value larger than 50 commuters between two towns.}
\label{fig:1}
\end{figure}

The standard mathematical representation of the resulting network is
provided by the adjacency matrix $A$ of elements ($a_{ij}$). The
elements on the principal diagonal ($a_{ii}$) are set equal to zero,
since intra-municipal commuting movements are not considered
here. Off-diagonal terms $a_{ij}$ are equal to $1$ in the presence of
any non-zero flow between $i$ and $j$ ($i\to j$ or $j\to i$) and are equal
to $0$ otherwise. The adjacency matrix is then symmetric and describes
regular bi-directional displacements among the municipalities.  The
adjacency matrix contains all the topological information about the
network but the dataset also provides the number of commuters attached
to each link. It is therefore possible to go beyond the mere
topological representation and to construct a weighted graph where the
nodes still represent the municipal centres but where the links are
valued according to the actual number of commuters. Analogously to the
adjacency matrix $A$, we thus construct the symmetric weighted
adjacency matrix $W$ in which the elements $w_{ij}$ are computed as
the sum of the $i\to j$ and $j\to i$ flows between the corresponding
municipalities (per day). The elements $w_{ij}$ are null in the case
of municipalities $i$ and $j$ which do not exchange commuting traffic
and by definition the diagonal elements are set to zero . According to
the assumption of regular bi-directional movements along the links,
the weight matrix is symmetric and the network is described as an
undirected weighted graph. The weighted graph provides a richer
description since it considers the topology along with the
quantitative information on the dynamics occurring in the whole
network. It is however important to stress that while the nodes
correspond to municipalities located in the physical space, the graph
representation does not contain any information explicitly related to
geographical distances and other spatial characteristic of the
network. The definition of a network representation that correlates
topological and traffic characteristics with the spatial properties of
the SMCN requires more refined investigation and will be presented
elsewhere (De Montis et al, forthcoming).

\section{The analysis of the topological properties of the SMCN}

In this section, we analyze the topology of the SMCN and we propose
possible territorial explanations for our findings. This network is
relatively small being characterized by $N=375$ vertices and
$E=16,248$ edges. The degree $k$ of nodes, which measures the number
of links of each node, ranges between $8$ and $279$ and exhibits a
high average value . The degree can be considered as a first
indication of the topological centrality of a vertex: the importance
of the corresponding municipal centre as attraction point for
commuters in the network. A measure of the topological distance within
the network is given by the shortest path length defining the distance
between nodes in terms of links to be traversed by using the shortest
possible path between the nodes. On the average, the shortest path
length is $l\approx 2.0$, while the maximum length is just equal to
$3$. These values are small compared to the number of nodes, in
agreement with a small-world behavior for which l typically scales as
the logarithm of $N$.  Further information on the network's topology
is provided by the degree distribution $P(k)$ defined as the
probability that any given node has degree $k$. In this case it is
essential to distinguish between an 'exponential' network similar to
the usual random graph for which $P(k)$ decreases exponentially fast,
and scale-free networks indicative of hub-like hierarchies for which
$P(k)$ decreases typically as a power law (Albert and Barabasi, 2002;
Amaral et al, 2000). In the present case, the degree probability
distribution is skewed and is relatively peaked around a mean value of
order $40$ (Figure 2).
\begin{figure}[ht]
\vskip .5cm
\begin{center}
\epsfig{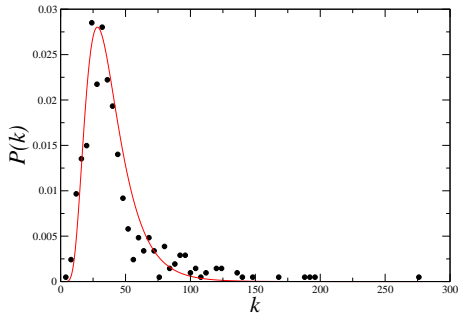}
\end{center}
\caption{Plot of the probability distribution of connectivity. The red line is a lognormal fit.}
\label{fig:2}
\end{figure}
The tail of the distribution contains outliers with respect to a
regular exponential decay, signaling the presence of municipal hubs,
which exchange commuters with many others municipalities. However, a
closer inspection of the distribution tail shows that the decay is not
compatible with a heavy-tailed distribution. The tail has indeed a
power-law behavior however with an exponent that ensures the
convergence of the relevant moments of the distribution and does not
imply a scale-free heterogeneity. In particular, the variance
converges to a finite limit for infinitely large systems in contrast
with scale-free networks, which have an infinite variance for infinite
size networks.

A further relevant topological quantity is the clustering coefficient
defining a measure of the level of cohesiveness around any given
node. It is expressed as the fraction of connected neighbors
\begin{equation}
C(i)=\frac{2E_i}{k_i(k_i-1)}
\end{equation}
where $E(i)$ is the number of links between the neighbors of the node
$i$ and $k_i(k_i-1)/2$ is the maximum number of possible
interconnections among the neighbors of the node. This quantity is
defined in the interval $[0,1]$ and measures the level of local
interconnectedness of the network. If $C(i) = 0$, the neighbors of the
node $i$ are not interconnected at all, while $C(i) = 1$ corresponds
to the case where all the neighbors are interconnected. A large
clustering thus indicates the existence of a locally well-connected
neighborhood of nodes. It is often convenient to average $C(i)$ over
all nodes with a given degree $k$ obtaining the clustering spectrum
\begin{equation}
C(k)=\frac{1}{NP(k)}\sum_{i/k_i=k}C(i)
\end{equation}
where $NP(k)$ is the total number of nodes of degree $k$. In Figure 3, on
the left, we plot the clustering spectrum of the SMCN: we observe a
decaying behavior with $C(k)$ that varies from $0.8$ to values lower than
$0.2$.
\begin{figure}[ht]
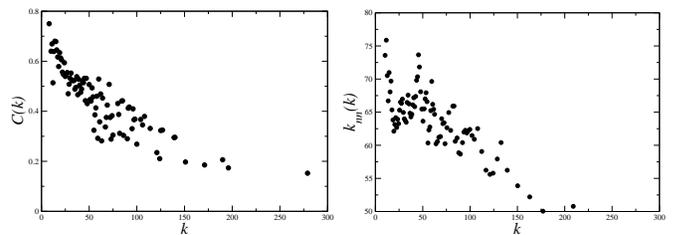

\vspace*{.1cm} 
\centerline{
\includegraphics*[width=0.24\textwidth]{./newC_k.eps}
\includegraphics*[width=0.24\textwidth]{./newknn_k.eps}
}
\caption{ On the left: Scatterplot of the clustering coefficient
versus degree. On the right: assortativity of the SMCN showing a
slight disassortative behaviour.}
\label{fig:3}
\end{figure}
This indication of a decreasing clustering suggests that the SMCN
behaves like other studied real networks, in that low degree
municipalities belong to well interconnected working or studying
communities while the high degree vertices, the urban hubs, connect
otherwise disconnected regions. It is worth noting that this result is
similar to the one obtained by Chowell et al (2003) for urban
movement. There is a striking difference however since the SMCN is
topologically very close to a random graph while the network studied
in Chowell et al (2003) is a scale-free network. This similarity with
a random graph is also confirmed by the average value of the
clustering coefficient $\langle C\rangle=0.26$, that results to be of
the same order of the clustering coefficient theoretically computed
for the case of a generalized random graph, which is
\begin{equation}
\langle C\rangle_{RG}=\frac{(\langle k^2\rangle-\langle k\rangle)^2}{N\langle k\rangle^3}
\end{equation}
predicting the value $\langle C\rangle_{RG}\approx 0.24$.

Another important property is the degree similarity of the neighbors
of a node. This is measured by the average degree of the nearest
neighbors of a given node $i$
\begin{equation}
k_{nn}(i)=\frac{1}{k_i}\sum_{j\in{\cal V}(i)}k_j
\end{equation}
where ${\cal V}(i)$ denotes the set of neighbors of $i$. Analogously to
the clustering coefficient, we can average the assortativity over
nodes with a given degree leading to
\begin{equation}
k_{nn}(k)=\frac{1}{NP(k)}\sum_{i/k_i=k}k_{nn}(i)
\end{equation}
This spectrum measures the tendency of vertices to be connected with
vertices with the same degree properties.  If $k_{nn}(k)$ increases with $k$, nodes
have a tendency to connect to nodes with a similar or larger degree
and the network is said to be assortative. In contrast, if $k_{nn}(k)$ is
decreasing with $k$, small degree nodes connect preferentially to hubs
and the network is defined as disassortative. The result obtained for
the SMCN is shown in Figure 3, on the right, and displays a $k_{nn}$ with
decreasing trend. This result seems to indicate that the SMCN belongs
to the class of disassortative mixed networks. The disassortative
behavior is typical of technological and transportation network and
signals the presence of a global hierarchy in which the hubs tend to
provide the connectivity to the small degree nodes at the periphery of
the network.

\section{The SMCN weighted network analysis}

The ODT dataset provides the number of commuters among municipalities,
allowing the construction of the weighted graph representation of the
SMCN. The analysis of these commuting flows informs us on the system
dynamical behavior and on the needs of the commuters in terms of
regional transportation services and infrastructure. This information
might provide relevant indication to researchers and decision-makers
for addressing environmental policies and planning.

\subsection{Weight analysis}

The values of the weights between pairs of vertices ranges between $1$
and $w_{max}=13,953$ with an average value $\langle w\rangle\approx
23$ much lower than $w_{max}$. This fact is a signature of a high
level of heterogeneity in the SMCN, since a peaked distribution of
weights would very unlikely give a maximum value that is 3 orders of
magnitude larger than the average. Indeed, the probability
distribution $P(w)$ that any edge has a weight $w$ displays a power-law
decay $P(w)\sim w^{-\gamma_w}$ with an exponent $\gamma_w\approx 1.8$ (Figure 4, on the left).
\begin{figure}[ht]
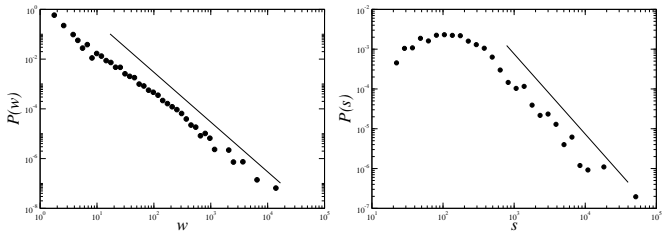

\vspace*{.1cm} 
\centerline{
\includegraphics*[width=0.24\textwidth]{./Pw.eps}
\includegraphics*[width=0.24\textwidth]{./Ps.eps}
}
\caption{On the left: Log-Log plot of the probability distribution of
the weights (the straight line is a power law fit of exponent $1.8$). On
the right: Log-Log plot of the probability distribution of the strength
(as a guide to the eye we also plot a power law fit of exponent $2$).}
\label{fig:4}
\end{figure}

Another relevant quantity that characterizes the traffic is given by
the strength of a node, defined as
\begin{equation}
s(i)=\sum_{j\in{\cal V}(i)}w_{ij}
\end{equation}
This quantity corresponds to the total traffic commuting in the town
$i$ and can be considered as a measure of the traffic centrality of a
municipality. For the SMCN, the strength varies between $1$ and
$64,834$ while its average value is equal to $1,010$. The probability
distribution of the strength $P(s)$ that any given node has a strength
$s$ is fitted by a power-law regime $P(s)\sim s^{-\gamma_s}$ with
exponent $\gamma_s\approx 2$ (Figure 4, on the right). The strength
provides an intuitive parameter for the description of the centrality
of Sardinian municipalities according to the actual traffic handled
generated and received by the municipality.

In the first instance, the variation in the weight and strength values
implies that commuters' flows are distributed in a broad spectrum
with a large heterogeneity. It is worth commenting the implications of
a power-law behavior of the statistical distributions $P(s)$ and
$P(w)$. This behavior defines the so-called heavy tailed distributions
which have a virtually unbounded variance. A peculiar fact about a
distribution with a heavy tail is that there is a finite probability
of finding vertices with weight or strength much larger than the
average value. In other words, the consequence of heavy tails is that
the average behavior of the system is not typical. The characteristic
weight (or strength) is the one that, picking up a vertex at random,
should be encountered most of the times. This is evident in the
bell-shaped distribution observed for the degree of the SMCN, in which
the average value is very close to the maximum of the distribution and
represents the most probable value in the system.  On the contrary,
the distributions shown in Figure 4 are highly skewed. Vertices have
an appreciable probability of having large strength values, yet all
intermediate values are probable and their mean does not represent any
special value for the distribution. The power-law behavior and the
relative exponent thus represent a quantitative measure of the level
of heterogeneity of the network's traffic.

Another important question concerns the distribution of the traffic
among the different roads.  All the connections could carry a similar
flow or on the contrary one connection could dominate. A convenient
measure of this is given by the disparity (Barth\'elemy et al, 2005),
defined as
\begin{equation}
Y_2(i)=\sum_{j\in{\cal V}(i)}\left(\frac{w_{ij}}{s_i}\right)^2
\end{equation}
For hubs ($k\gg 1$), this quantity enables to distinguish situations
(Figure 5, on the right) for which all weights are of the same order
($Y_2\approx 1/k\ll 1$) from situations where only a small number of connections
dominate ($Y_2$ is of order 1/n where $n\ll k$).

We computed the value of $Y_2$ for each node and we average these
quantities for each degree in order to obtain $Y_2(k)$. The result is
shown on Figure 5, on the left. We obtain here an average behavior of
the form $kY_2(k)\sim k^{1-\theta}$ with $\theta\approx 0.4$. This
result shows that the weights on the links attached to a given node
are very heterogeneous. In other words, there are only a few dominant
connections, while the traffic on all the other roads is very small.
\begin{figure}[ht]
\vspace*{.1cm} 
\centerline{
\includegraphics*[width=0.25\textwidth]{./newY2_k.eps}
\includegraphics*[width=0.25\textwidth]{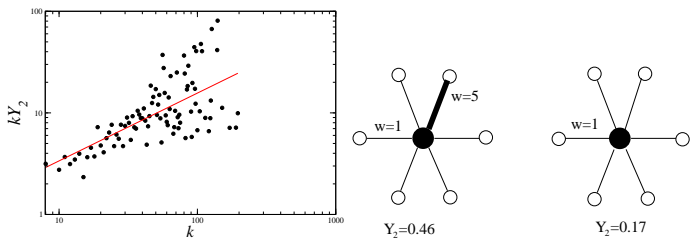}
}
\caption{On the left: Log-Log plot of the disparity versus connectivity for the
SMCN; on the right: illustration of the disparity for two very
different cases. When a few connections dominate, $Y_2$ is of order $1$; in
contrast if all connections have the same weight, $Y_2$ is of order $1/k$.}
\label{fig:5}
\end{figure}

\subsection{The analysis of traffic-topology correlations}

In order to inspect in more detail the relation between the degree and
the traffic, we show in Figure 6 the strength of nodes as a function
of their degree. Despite the existence of inevitable fluctuations, we
observe over a wide range of degrees a power-law behavior of the form
$s(k)\sim k^{\beta}$ with an exponent $\beta\approx 1.9$. Independent
weights and connectivities would give a value $\beta=1$ (Barrat et al, 2004)
and the result obtained here reveals that there is strong correlation
between the traffic and the topology. The strength of the nodes grows
faster than their degree: the more a municipality is connected with
other centres the much more it is able to exchange commuters' flows
or, in other terms, the traffic per connection is not constant and
increases with the number of connections.
\begin{figure}[ht]
\epsfxsize=2.0in
\centerline{\epsffile{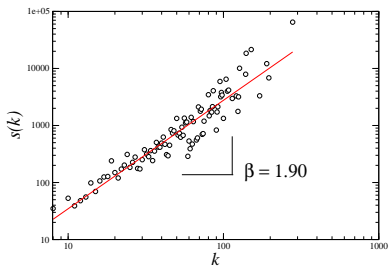}}
\vspace*{0.3cm}
\caption{ Average strength of the municipalities as a function of the degree. }
\label{fig:6}
\end{figure}

The presence of weights also alters the definition of other structural
measures. For instance, the standard topological clustering
coefficient used in the previous section needs to be augmented by the
inclusion of the actual traffic. In order to understand the weight
structure and their relation with topology, the weighted clustering
coefficient can be defined as follows (Barrat et al, 2004)
\begin{equation}
C^w(i)=\frac{1}{s_i(k_i-1)}\sum_{j,h}\frac{w_{ij}+w_{ih}}{2}a_{ij}a_{ih}a_{jh}
\end{equation}
where $a_{lm}$ is an element of the adjacency matrix and where
$s_i(k_i-1)$ is a normalization factor which ensures that $C^w(i)$
belongs to $[0,1]$. As for the topological case it is possible to
average over all nodes of same degree $k$ obtaining the clustering
spectrum $C^w(k)$. The weighted clustering coefficient counts for each triple
formed in the neighborhood of the vertex i the weight of the two
participating edges starting from i. In this way, we are not just
considering the number of closed triangles but also their total
relative weight with respect to the vertex' strength. In the case of
random networks $C^w=C$ but in real weighted networks, we can however face two
different situations. If $C^w<C$ the topological clustering is generated by
edges with low weight and therefore the cohesiveness is less important
in terms of traffic properties. On the contrary, if $C^w>C$ we are in presence
of a network in which the interconnected triples are more likely
formed by edges with larger weight. We show in Figure 7 (on the right)
an example of such a situation. It is clear that the interconnected
triple has a major role in the network dynamics and organization, and
that the clustering properties are not completely exploited by a
simple topological analysis. The weighted clustering coefficient takes
into account the weighted elements by an appropriate mathematical
form.

In the SMCN the values of the weighted clustering coefficient are
larger than the corresponding topological values $C^w>C$ (Figure 7, on the
left) over the entire range of degree values. In contrast with the
topological clustering, the weighted clustering coefficient is
approximately constant over the whole range of connectivity.
\begin{figure}[ht]
\vspace*{.1cm} 
\centerline{
\includegraphics*[width=0.27\textwidth]{./newfigure7.eps}
\includegraphics*[width=0.20\textwidth]{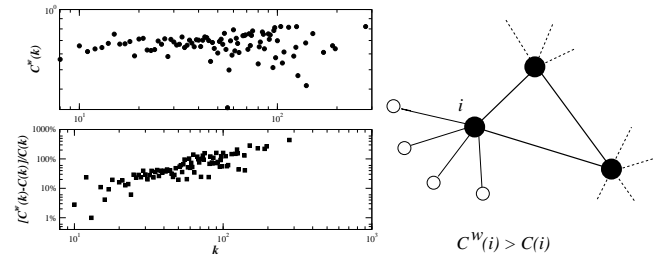}
}
\caption{ On the left, top: weighted clustering coefficient versus
connectivity; on the left, bottom: relative difference between the
weighted and the topological clustering coefficients versus
connectivity; on the right: large cities are well interconnected by
large flow links (bold edges must be considered with a larger weight).}
\label{fig:7}
\end{figure}
This result reveals the existence of a rich-club phenomenon in that
important cities form a group of well interconnected nodes and that
weight heterogeneity is enough in order to balance the lack of
topological clustering. The rich-club phenomenon implies that in the
SMCN the interconnected triplets are more frequently built by edges
with a higher weight (Figure 7, on the right): in terms of local
cohesiveness of workers' and students' commuting system, two
different destinations available from a given city are more likely to
be connected if the traffic flow leading to them is large.  Similarly
to the weighted clustering, we need to integrate the information on
weights in a sensible definition of
assortativity/disassortativity. This can be easily done by introducing
the weighted average degree of the nearest neighbors of a given node $i$
\begin{equation}
k^{w}_nn(i)=\frac{1}{s_i}\sum_{j\in{\cal V}(i)}w_{ij}k_j
\end{equation}

\begin{figure}[ht]
\vspace*{.1cm} 
\centerline{
\includegraphics*[width=0.27\textwidth]{./newfigure8.eps}
\includegraphics*[width=0.20\textwidth]{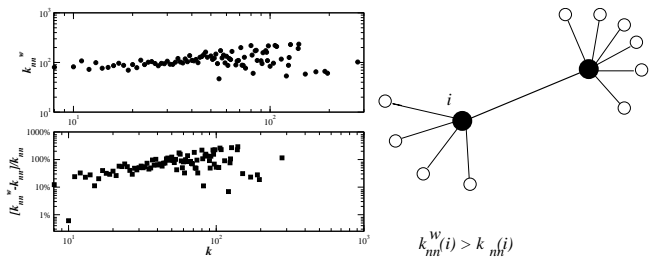}
}
\caption{ On the left, top: weighted assortativity; on the left,
bottom: relative difference between the weighted and the topological
assortativity; on the right: a difference between the weighted and the
topological assortativity indicates that large cities are connected by
large flow links.}
\label{fig:8}
\end{figure}
In this case, we perform a local weighted average of the nearest
neighbor degree according to the normalized weight of the connecting
edges. This definition implies that $k^{w}_{nn}> k_{nn}$ if the edges
with the larger weights are pointing to the neighbors with larger
degree and $k^{w}_{nn}< k_{nn}$ in the opposite case. The $k^{w}_{nn}$
thus measures effective affinity to connect with high or low degree
neighbors according to the magnitude of the actual interactions. In
the SMCN the weighted average degree of the nearest neighbors displays
a slightly increasing pattern, which is a signature of assortativity
(Figure 8). This result is in sharp contrast with the behavior of the
topological disassortativity and indicates an affinity of high-degree
municipalities with other large centers with which they exchange high
number of commuters: hubs have more frequently large traffic flows
with municipalities with the same or higher-order degree.  This
phenomenon is illustrated on Figure 8, on the right: large weights
connect hubs while the connection with nodes of smaller degree has
smaller weights.

\section{Discussion: the SMCN and its relation with real environmental systems}

\subsection{Topology and environmental properties}

The ranking connected to the degree (Table 1), an index of topological
centrality, confirms the existence of administrative and
socio-economic hierarchies: among the hubs, Cagliari is the most
populated and regional capital town, Nuoro, Oristano and Sassari are
province capital towns, Macomer and Quartu Sant'Elena are emerging
productive and residential centres.
\begin{table}
\begin{center}
\begin{tabular}{|c|c|c|}
\hline
Rank  & Municipal centers & Degree $k$ \\
\hline
\hline
$1$ & Cagliari &  279 \\
\hline
$2$ & Nuoro & 196  \\
\hline
$3$ & Oristano &  190 \\
\hline
$4$ & Macomer &   171\\
\hline
$5$ & Sassari &   151\\
\hline
$6$ & Quartu Sant'Elena & 140  \\
\hline
$7$ & Assemini & 139  \\
\hline
$8$ & Selargius &  127 \\
\hline
$9$ & Villacedro &  125 \\
\hline
$10$ & Ottana &  124 \\
\hline
\end{tabular}
\caption{Ranking of Sardinian municipalities by their degree $k$.}
\label{tab:1}
\end{center}
\end{table}
While the SMCN network is not scale free, the analysis of the degree
probability distribution still reveals the presence of a relatively
large number of well-connected towns that attract commuters from a
fairly high number of satellite centers. The behavior of the
clustering coefficient versus the degree indicates that for $k< 50$ the
clustering is relatively high and is a signature of the fact that
small towns (small $k$) form very well connected clusters of
municipalities exchanging students and workers. On the contrary,
hub-towns ($k >50$) have less cohesive neighborood since they have
commuting flows with several different local clusters of
municipalities not interconnected among them. This kind of star-like
topology with large municipalities acting as hub for several smaller
groups of municipalities is confirmed by the analysis of the average
nearest neighbor degree spectrum suggesting that the SMCN belongs to
the class of dissassortative mixed networks. In this case, high degree
hubs link preferentially with low degree municipalities. This
topological analysis would suggest the existence of inter-municipal
commuting districts of small towns, which pivot around a few urban
poles and could be interpreted as the signature of a phenomenon often
observed in urban and transportation systems when top functional rank
towns attract commuters from small centers, behaving as urban
poles. In such an environment, citizens of the satellite centers,
being interested to regional-level services such as public
administration, health and finance, prefer commuting to the pole
rather than to a peer level town. The heterogeneity detected in the
pattern of municipal centers could be seen as functional to an
efficient behavior of the whole network, since each urban hub exchange
commuters, and thus goods, services and revenues, with several small
size towns. Turning to the geographical displacement of the hub
centers, we can note that as expected they are located far away from
each other. The distance between Cagliari and Sassari is equal to $210$
Km, while from Cagliari and Nuoro to $160$ Km, with respect to a total
length of the island equal to $230$ Km.

\subsection{Traffic and environment in the weighted characterization of the SMCN}

 The heterogeneity of the weights, as we saw in the previous chapters,
 is very high. This implies that the links are not all equivalent and
 thus warns against any interpretation based on the sole
 topology. Indeed, the analysis of the weighted network introduces new
 insights in the interpretation elaborated on the result of the
 topological properties under a variety of perspectives.

\begin{table}[ht]
\begin{center}
\begin{tabular}{|c|c|c|}
\hline
Rank  & Pairs of connected municipal centers & Weight $w$ \\
\hline
\hline
$1$ & Cagliari-Sassari &  13,953 \\
\hline
$2$ & Sassari-Olbia & 7,246  \\
\hline
$3$ & Cagliari-Assemini &  4,226 \\
\hline
$4$ & Porto Torres-Sassari &  3,993 \\
\hline
$5$ & Cagliari-Capoterra &   3,731 \\
\hline
\end{tabular}
\caption{Ranking of pairs of Sardinian municipal centres by the weight of their connections.}
\label{tab:2}
\end{center}
\end{table}

Since the weights are distributed in a very heterogeneous pattern, the
corresponding commuters' flows display a wide range of values, which
reflects the existence of a territorial hierarchy embedded and
supported by the underlined transportation network. Precise
indications emerge about dominant relations between some of the major
poles of the island (Table 2): Cagliari, Sassari, Olbia, Assemini,
Porto Torres and Capoterra.

In particular, the analysis of the strength probability distribution
shows that municipalities identified as topological poles behave also
as attractors of traffic flows. The strength is broadly distributed
along the nodes, revealing a strong hierarchical dominance of the
dynamic hubs over other smaller level centers.
\begin{table}[ht]
\begin{center}
\begin{tabular}{|c|c|c|}
\hline
Rank  & Municipal centers & Strength $s$ \\
\hline
\hline
$1$ & Cagliari &  64,834 \\
\hline
$2$ & Sassari & 21,437 \\
\hline
$3$ & Quartu Sant'Elena &  18,431 \\
\hline
$4$ & Oristano &  12,130 \\
\hline
$5$ & Selargius &   10,084 \\
\hline
$6$ & Assemini &   7,915 \\
\hline
$7$ & Porto Torres &  6,886 \\
\hline
$8$ & Nuoro &   6,834 \\
\hline
$9$ & Carbonia &   6,616 \\
\hline
$10$ & Iglesias & 6,479 \\
\hline
\end{tabular}
\caption{Ranking of Sardinian municipalities by their strength.}
\label{tab:3}
\end{center}
\end{table}
The analysis of the relation between strength and degree shows that
traffic grows superlinearly with the degree confirming the fact that
the more the topological hub-towns are connected and the larger their
traffic centralities (ie. their capacity to attract Sardinian
commuting students and workers). The ranking of Sardinian
municipalities according to their strength is reported in Table 3.

From the behavior obtained for the disparity (Fig. 5), we see that
only few connections emerging from the hub-towns carry a large amount
of traffic, indicating a strong hierarchy and heterogeneity also in
the local structure of the flows connecting large municipalities of
the SMCN. This mirrors the underlying road network structure, which
comprises a very small set of main highways (state roads), a large set
of medium-sized roads (provincial roads) and a very large set of local
roads (municipal roads). In particular, the link Cagliari-Sassari
corresponds to state road number $131$, named 'Carlo Felice' that is the
highway crossing longitudinally the island and can be considered as
the backbone of the SMCN structure to which the lower hiearchical
levels transportation structures connect.

The use of the traffic information provides further insight on the
structure and hierarchy of the network. In particular the difference
detected between weighted and topological clustering coefficients and
weighted versus topological assortativity indicates the presence of a
distribution of traffic flows that might be indicative of what
sometimes is referred to as the rich-club phenomenon. Indeed, it is
possible to observe that large traffic flows are characterizing the
commuter traffic among municipalities with high centrality ranks as
measured in terms of degree and strength. The accumulation of weight
on these centers attenuates the decaying of the clustering and average
nearest neighbour spectrum that appears to invert their behaviour by
using the weighted definitions. This is the sign of the existence of a
backbone of commuting patterns among large municipalities which are
all interconnected, while the flows toward peripheral municipalities
are indeed of small size.  According to this picture, small
municipalities are mostly connected each other through small traffic
links, which correspond to the underlying second or third order
roads. This completes the topological information and offers a clear
picture of the ordering principles of the SMCN, in which a large
number of small-size cities tend to become satellites of higher rank
municipalities, shaping an overall network structure widely punctuated
with star-like subsystems pivoting around important urban poles
connected through dominant roads and the corresponding traffic
flows. In this respect it appears that the network analysis presented
here is able to provide an extremely detailed and quantitative
characterization of the structure of the inter-urban traffic system.

\subsection{Relating topology and traffic to socio-economic phenomena}

In this section, we provide a discussion of the emergent topological
and traffic properties in relations to the demographic and economic
properties of the towns of the SMCN.
\begin{figure}[ht]
\vspace*{.1cm} 
\centerline{
\includegraphics*[width=0.20\textwidth]{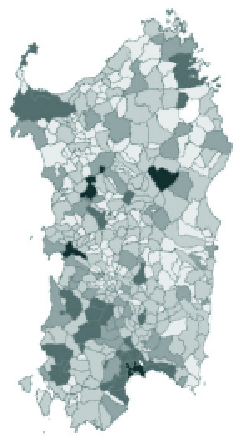}
\includegraphics*[width=0.20\textwidth]{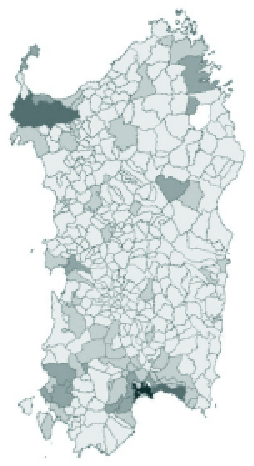}
}
\caption{ On the left: map of the degree k; on the right: map of the
strength s for each Sardinian municipality (Scale: the darker the
shade, the higher the value).}
\label{fig:9}
\end{figure}
In the previous sections, we observe that the values of degree k and
of the strength s, as indexes of topological and weighted centrality,
confirm quantitatively historical structural hierarchies among
Sardinian towns. The heterogeneity of the strength and the degree can
be visualized as shown in the thematic map representations of Figure
9.

We explore now in quantitative terms the relation between the network
properties and the environmental and economical indicators. One goal
is to inspect whether the commuting network hubs correspond to
economically relevant towns. As a measure of social and economic
centrality of each town, we consider two variables: the total resident
population (pop) in 1991 (Istat, 1991b) and the average monthly income
(mmi). The average monthly income, an index of urban aggregated
wealth, is defined as the product of the average monthly income per
worker times the number of workers (Carcangiu et al, 1993). A map
representation using these indicators is shown in Figure 10.
\begin{figure}[ht]
\vspace*{.1cm} 
\centerline{
\includegraphics*[width=0.20\textwidth]{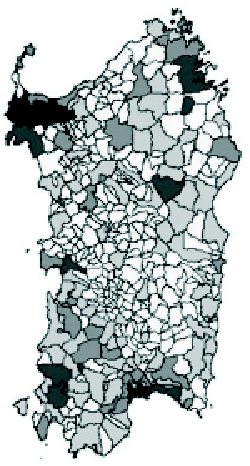}
\includegraphics*[width=0.20\textwidth]{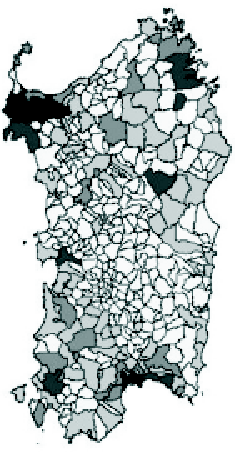}
}
\caption{ On the left: map of the resident population pop; on the
right: map of the municipal monthly income mmi for each Sardinian
municipality in 1991 (Scale: the darker the shade, the higher the
value).}
\label{fig:10}
\end{figure}

In Figure 11, we plot the variable $pop$ as a function of the degree $k$
and the strength $s$ and we observe a clear positive correlation. The
association between the population and these variables has a power-law
behavior with two different exponents related by a factor $1.9$. This is
expected since we have already observed that traffic and degree are
related by an empirical law $s~k^{1.9}$. This implies that if the
population has a scaling $pop~s^a$, the simple substitution of variable
yields $pop~k^{1.9a}$. This is confirmed in the analysis where the slopes
of the two curves are related by a factor close to two. It is
interesting to note that the population scales almost linearly with
the traffic which suggest that a constant fraction of the population
of every city commutes to another town, irrespectively of the size of
the city.
\begin{figure}[ht]
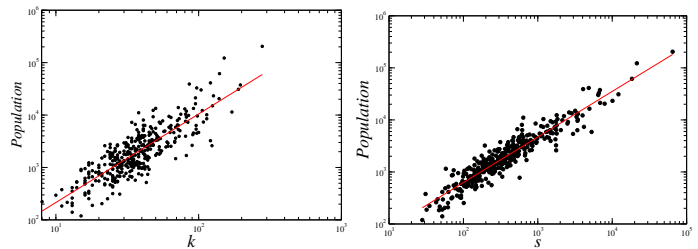

\vspace*{.1cm} 
\centerline{
\includegraphics*[width=0.25\textwidth]{./pop.vs.k.eps}
\includegraphics*[width=0.25\textwidth]{./pop.vs.s.eps}
}
\caption{ On the left: Log-Log plot of population over the degree $k$,
slope coefficient equal to $1.70$; on the right: Log-Log plot of
population over the strength $s$, slope coefficient equal to $0.90$.}
\label{fig:11}
\end{figure}

In Figure 12, the monthly municipal income ($mmi$) is plotted versus the
degree $k$ and the strength $s$. A positive correlation is signaled by a
linear behavior on a log-Log scale which indicates the presence of a
power law association. Also in this case, the exponents of the two
relations confirm the general result of a quadratic relation between
strength and degree.
\begin{figure}[ht]
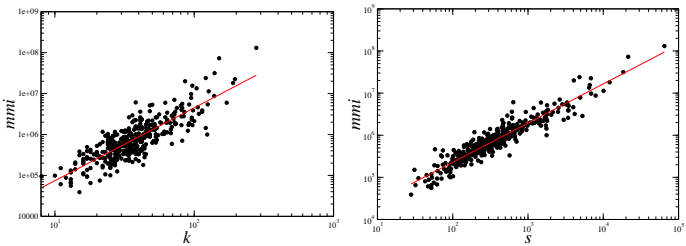

\vspace*{.1cm} 
\centerline{
\includegraphics*[width=0.25\textwidth]{./mmi.vs.k.eps}
\includegraphics*[width=0.25\textwidth]{./mmi.vs.s.eps}
}
\caption{ On the left: Log-Log plot of $mmi$ versus the degree $k$, the slope
coefficient is equal to $1.80$; on the right: Log-Log plot of $mmi$ versus the
strength $s$, the slope coefficient is equal to $0.90$.}
\label{fig:12}
\end{figure}

The results of this section confirm that the centrality measures
(strength and degree) display trends consistent with the behavior of
social and economic indicators such as population and income. In this
sense, the higher the network centrality of a node, the higher is its
demographic and economic size. The aforementioned territorial size
measures, $pop$ and $mmi$ grow much faster with the degree $k$ than
with the strength $s$.  We can therefore identify a group of variables
related to population ($s$, $pop$, $mmi$) that are correlated by
association laws with a behavior very close to the linear one. Their
relation with the degree is however highly non-linear indicating that
these variables grows much faster than the connectivity of
municipalities. Indeed, inverting the relation between degree and
traffic (or population) we have $k\sim\sqrt{s}$. The degree of a
municipality is both an indicator of the level of economic and social
exchange with other municipality and the transport infrastructure. The
above relation implies that the economic and transportation
infrastructures seem to grow at a pace slower than population
indicators. It is not clear if this is a peculiar characteristic of
the Sardinia region, due to political and urban planning decision, or
a general property of population flows indicating a kind of adaptive
equilibrium of a societal system. It is interesting to note that the
worldwide airport network exhibits association trends among the
network infrastructure and the population size very similar to those
observed in this study (Barrat et al, 2004), suggesting a possible
level of universality. The detailed study of other commuting and
transportation networks is however needed in order to gain further
insight on this issue.

\section{Conclusions}

In this paper, the inter-urban commuting system of the island of
Sardinia has been represented as a weighted network. This
representation allows the quantitative characterization of the
capacity of attraction of each urban center on workers and
students. By comparing the different levels of hierarchies emerging in
the network with the pattern of demographic and economic poles we shed
light on several structural and organizational properties of the
inter-municipal traffic system.

At first sight, the SMCN is similar to a small-world random (Watts and
Strogatz, 1998). The clustering organization of the SMCN however
diverges from the usual random paradigm and similarly to other
technological networks, such as the airline network, suggests a
hierarchy of nodes where small municipalities are locally densely
interconnected, while large municipalities (regional hub) provide the
long range connectivity for different areas of the islands otherwise
disconnected. This picture is reinforced by the evidence of a
disassortative mixed network where the hub municipalities are found to
act as star-like vertices connecting on average municipalities with
low degree and low centrality ranking.

The weighted representation of the network allowed by the inclusion of
traffic data complements the resulting picture with important
information relating the commuter traffic to the available topological
connectivity. We find that both the weights and the strengths are very
broadly distributed, confirming the necessity of including those
measures in a realistic description of the SMCN. In particular, the
probability distributions of weights and strengths (commuter traffic
handled by the municipality) display a power-law regime over a wide
range of degree values: the SMCN behaves as a scale-free weighted
network. In addition, the traffic is strongly correlated to the
connectivity of the municipality and moreover, as shown by the
disparity, it is largely concentrated on a very few links. This
indicates that the more a town is connected, the much more it attracts
commuters along a small number of dominant roads. Our analysis of
weighted clustering suggests that important towns form a group of
well-interconnected centers. It also shows that in the SMCN the main
fraction of traffic exchanges occurs among the hubs of the network:
hub centers accumulate their commuter traffic to same or higher order
municipalities. In terms of environmental system analysis, a large
number of small-size centers are commuting satellites of higher rank
municipalities. The SMCN appears as widely punctuated with star-like
subsystems pivoting around important urban poles connected
preferentially through high traffic roads. In addition, the fact that
the weighted clustering coefficient is large for the whole range of
connectivity suggests that on the local scale there are
micro-commuting basins, rationalized as inter-municipal districts of
highly clustered small towns, which pivot around a few urban poles
belonging to macro commuting basins.

Interestingly enough, there are striking similarities for the weight
distribution and the clustering obtained in the present study with
those obtained for the networks of population movements at the urban
level (Chowell et al, 2003): this evidence suggests the existence of a
set of basic mechanisms leading to common features in transportation
systems at various granularity levels. In Table 4, the main
characteristics of different transportations networks at different
scales are shown. In particular, it seems that while the topological
properties can vary, the broadness of the traffic distribution is a
common feature to all these networks. Another common feature is the
existence of correlation between the traffic and the topology.

\begin{table}[ht]
\begin{center}
\begin{tabular}{|c|c|c|c|c|}
\hline
Network & $P(k)$  & $P(s)$ & $s\sim k^\beta$  & $Y_2\sim k^{-\theta}$ \\
\hline
\hline
Global: WAN                 & Heavy tail  & Broad &  $\beta=1.5$ & $\theta=1.0$\\
\hline
Inter-Cities                      & Light tail  & Broad &  $\beta=1.9$ & $\theta=0.4$\\
\hline
Intra-urban   & Heavy tail  & Broad &  $\beta=1.0$ & n/a \\
\hline
\end{tabular}
\caption{Comparison of properties of transportation networks.}
\label{tab:4}
\end{center}
\end{table}

Finally, the existence of the municipality hierarchy that we have
uncovered with the help of a weighted network analysis is confirmed by
comparing demographic and economic urban polarization in
Sardinia. Population and wealth, as proxies of the endowment of local
resources, display a positive correlation with topological and dynamic
centrality, strengthening the value of the network approach as a
general tool for the analysis of inter-urban transportation systems.

\section{Outlook}

The results of this investigation open further questions that are
currently under study. The indication of network centrality provides a
powerful measure of similarity of the municipalities in attracting
workers and students daily, more work however is needed to tackle the
problem of the precise identification of the communities of
workers/students commuting in Sardinia and their geographical
basins. These studies may provide insights for bottom-up processes of
decision-making and planning based on the definition of the existing
emerging behaviors of the citizens.  Finally, the SMCN is obviously
not independent from geography and space. It refers to a physical
network, mostly based on road systems, which display precise
geographical characteristics. Further studies which include spatial
aspects of the network are currently under progress. This forthcoming
analysis will hopefully provide decision-makers and planners with
relevant indications on the global effect of new infrastructure.

\begin{acknowledgments}

A.V. is partially funded by the European commission under the project
contract 001907 (DELIS). M.B. is on leave of absence from CEA-Centre
d'Etudes de Bruy\`eres-le-Ch\^atel, D\'epartement de Physique
Th\'eorique et Appliqu\'ee BP12, 91680 Bruy\`res-le-Ch\^tel,
France. The authors wish to thank Michele Campagna for his useful
comments and discussions.

\end{acknowledgments}

\section{References}

\begin{enumerate}

\item{} Albert R, Barab\`asi AL, 2002, `Statistical mechanics of complex
networks', Rev. Mod. Phys. {\bf 74}, 47-97.

\item{} Amaral LAN, Scala A, Barth\'elemy M, Stanley HE, 2000, `Classes of
small-world networks', Proc. Natl. Acad. Sci. (USA) {\bf 97},
11149-11152.

\item{} Barab\`asi AL, Albert R, 1999, `Emergence of scaling in random networks',
Science {\bf 286}, 509-512.

\item{} Barrat A, Barth\'elemy M, Pastor-Satorras R, Vespignani A, 2004, `The
architecture of complex weighted networks' Proceedings of The
National Academy of Sciences {\bf 11}, 3747-3752.

\item{} Barth\'elemy M, Gondran B, Guichard E, 2003, `Spatial Structure of the
Internet Traffic' Physica A {\bf 319}, 633-642.

\item{} Barth\'elemy M, Barrat A, Pastor-Satorras R, Vespignani V, 2005, 
`Characterization and modelling of weighted networks' Physica A {\bf 346},
34-43.

\item{} Batty M, 2001, `Cities as small worlds', Editorial Environment and
Planning B: Planning and Design {\bf 28}, 637-638.

\item{} Batty M, 2003, `Network geography' Paper 63, Centre of Advanced
Spatial Analysis (CASA), UCL.

\item{} Carcangiu R, Sistu G, Usai S, 1999, `Struttura socio-economica dei
comuni della Sardegna. Suggerimenti da un analisi cluster' 
(Socio-economic structure of Sardinian municipalities. Evidence from a
cluster analysis), Working Papers, {\bf 3}, CRENOS, Cagliari.

\item{} Chowell G, Hyman JM, Eubank S and Castillo-Chavez C, 2003, `Scaling
laws for the movement of people between locations in a large city' 
Physical Review E {\bf 68}, 066102.

\item{} De Montis A, Barth\'elemy M, Campagna M, Chessa A, Vespignani V,
forthcoming, `Topological and dynamic properties emerging in a real
inter-municipal commuting network: perspectives for policy-making and
planning', Congress of the European Regional Science Association,
Amsterdam, 23-27 August 2005.

\item{} Dorogovtsev SN, Mendes JFF, 2003, `Evolution of networks: From
Biological nets to the Internet and WWW' Oxford Univ. Press, Oxford.

\item{} Erd\"os P, Renyi P, 1960, `On the evolution of random graphs' 
Publ. Math. Inst. Hung. Acad. {\bf 5}, 17-60.

\item{} Gastner MT, Newman MEJ, 2004, `The spatial structure of networks' Cond-mat 0407680.

\item{} Gorman SP, 2001, `The network advantage of regions: the case of the
USA, Europe and China' paper presented at the International Regional
Science Association Meeting, Charleston, SC.

\item{} Gorman SP, Kulkarni R, 2004, `Spatial small worlds: new geographic
patterns for an information economy' Environment and Planning B:
Planning and Design {\bf 31}, 273-296.

\item{} Guimera R, Mossa S, Turtschi A, Amaral LAN, 2005, `The worldwide air
transportation network: Anomalous centrality, community structure, and
cities' global roles' Proc. Nat. Acad. Sci. (USA) {\bf 102},
7794-7799.

\item{} Italian National Institute of Statistics (Istat), 1991a, 13th Censimento
generale della popolazione e delle abitazioni, Matrice origine destinazione
degli spostamenti pendolari della Sardegna ({\bf 13}th General census of
population and houses, Origin destination matrix of the commuting movements
of Sardinia).

\item{} Italian National Institute of Statistics (Istat), 1991b, 13th Censimento
generale della popolazione e delle abitazioni, (13th General census of
population and houses).

\item{} Jiang B, Claramount C, 2004, `Topological analysis of urban street
networks' Environment and Planning B: Planning and Design {\bf 31} 151-162.

\item{} Latora V, Marchiori M, 2000, `Harmony in the small-world' Physica A {\bf 285}, 539-546.

\item{} Latora V, Marchiori M, 2001, `Efficient behaviour of Small-World
Networks' Phys. Rev. Lett. {\bf 87}, 198701.

\item{} Latora V, Marchiori M, 2002, `Is the Boston subway a small-world network?' Physica A {\bf 314}, 109-113.

\item{} Latora V, Marchiori M, 2003, `Economic small-world behavior in
weighted networks' The European Physical Journal B {\bf 32}, 249-263.

\item{} Newman MEJ, 2003, `The structure and function of complex networks' SIAM 
review {\bf 45}, 167-256.

\item{} Pastor-Satorras R, Vespignani A, 2004, `Evolution and Structure of
the Internet' Cambridge University Press, Cambridge, USA.

\item{} Salingaros NA, 2001, `Remarks on a city's composition' Resource for Urban Design Information.

\item{} Salingaros NA, 2003, `Connecting the Fractal City' Keynote speech,
5th Biennial of town planners in Europe, Barcelona.

\item{} Schintler AL, Gorman SP, Reggiani A, Patuelli R, Gillespie A, Nijkamp
P, Rutherford J, forthcoming, `Scale-free phenomena in
telecommunications networks' Networks and Spatial Economics.

\item{} Sen P, Dasgupta S, Chatterjee A, Sreeram PA, Mukherjee G and Manna SS,
2003, `Small World properties of the Indian Railway network' 
Phys. Rev. E {\bf 67}, 036106.

\item{} Shiode N, Batty M, 2000, `Power law distribution in real and virtual
worlds' Paper 19, Centre of Advanced Spatial Analysis (CASA), UCL.

\item{} Watts DJ, Strogatz SH, 1998, `Collective dynamics of 'small-world' 
networks' Nature {\bf 393}, 440-442.

\item{} Yook SH, Jeong H, Barab\`asi AL, 2002, `Modeling the Internet's
large-scale topology' Proc. Natl. Acad. Sci. (USA) {\bf 99}, 13382-13386.

\end{enumerate}


\end{document}